\documentclass{iopconfser}
\usepackage{ragged2e}
\usepackage{lineno}
\usepackage{xspace}
\usepackage{graphicx}
\usepackage{hyperref}
\hypersetup{
  hidelinks, 
  urlbordercolor = 1 1 1,
}

\def\Offline{\mbox{$\overline{\textrm%
{Off}}$\hspace{.05em}\protect\raisebox{.4ex}%
{$\protect\underline{\textrm{line}}$}}\xspace}

\begin{document}

\title{Auger Open Data and the Pierre Auger Observatory International Masterclasses}

\author{E Santos$^{1}$, for the Pierre Auger Collaboration$^{2}$}

\affil{$^1$Institute of Physics of the Czech Academy of Sciences, Prague, Czech Republic\\\vspace{0.55cm}}
\affil{$^2$Observatorio Pierre Auger, Av. San Mart\'{\i}n Norte 304, 5613 Malarg\"{u}e, Argentina\\
\vspace{0.01cm}
Full author list: \url{http://www.auger.org/archive/authors_2024_06.html}\vspace{0.15cm}}

\email{spokespersons@auger.org\vspace{0.1cm}}

\justifying
\begin{abstract}
The Pierre Auger Observatory has a public data policy following the FAIR principles
(Findable, Accessible, Interoperable, and Reusable).
We aim to share the data with the scientific community as part of the multi-messenger
effort at different levels and for educational activities to engage the general public.
Following the first portal created in 2007, a new portal hosted at
\url{https://opendata.auger.org} was established in February 2021.
The portal is regularly updated and comprises 10\% of the recorded cosmic-ray data
organized in various datasets, each with a specific DOI provided by Zenodo.
Moreover, a catalog with the 100 most energetic events is available.
The portal adopts a ``dual'' concept, offering not only the download of public data
but also a series of Jupyter notebooks.
These notebooks allow the general public to reproduce some of the most important
results obtained by the Pierre Auger Collaboration and understand the main mechanisms
governing the development of the extensive air showers produced by the interaction of
cosmic rays in the Earth's atmosphere.
In 2023, the Pierre Auger Observatory joined the International Particle Physics
Outreach Group (IPPOG).
The successful debut enrolled 550 high-school students at 12 research institutions
from 5 countries and was repeated this year, embracing yet more students and countries
worldwide.
During this day, the participants attend seminars about cosmic rays and are asked to
reconstruct subsets of public data events using an Auger 3-D event display.
Finally, they participate in a Zoom session with scientists at the Auger site.
\end{abstract}

\section{Introduction}

The Pierre Auger Collaboration has a long tradition in Outreach and Education that dates back
to its origin.
Over the more than two decades since the Pierre Auger Observatory was founded in the Argentinian
city of Malarg\"{u}e, the number of initiatives and ways of reaching out to the general public
has grown enormously.
The Visitor Centre in Malarg\"{u}e receives an annual average of around 8,000 people.
Every November, during the Pierre Auger Collaboration meeting, we organize a Science Fair where
school students from various educational levels from all over Argentina present their science
projects, and our collaborators participate as judges~\cite{karen_icrc2021}.
On our general website,~\url{https://www.auger.org/}, we have an Outreach tab dedicated to all
our activities, comprising the celebration of several international days, such as the
International Day of Women and Girls in Science, the Pi Day, the Hypathia Day (Equinox Day) or the
Earth hour, aimed at a broad audience ranging from other scientists to citizen scientists and the
general public of all ages.
We are also active on various social networks, where about 7,000 followers can be informed about our
latest news and projects.
Several activities are carried out during the year in schools and universities across several countries.
Our most recent initiative is the Auger International Masterclasses~\cite{raul_icrc2023}, in which
high-school students worldwide analyze the Auger public data, comprising 10\% of our
top-level cosmic-ray data used for our scientific publications, available at the Open Data portal hosted
at~\url{https://opendata.auger.org}, \cite{viviana_icrc2021, piera_icrc2023}.
In the following sections, the Auger Open Data portal and the activities performed
at the Pierre Auger International Masterclasses are described in detail.

\section{Open Data portal}

The Auger Open Data portal~\cite{open_data, PierreAuger:2023mbm}, hosted
at~\url{https://opendata.auger.org}, was established in February 2021.
One of its objectives was to widen the scope of the old data portal created in 2007.
During ten years, the old portal was updated every year with 1\% of the cosmic-ray data with
energy above $1\:\mathrm{EeV}$ detected by the surface detector array, with one year of latency,
for outreach and educational purposes.

The new Auger Open Data portal has a dual concept, allowing the Auger public data to be used for
scientific research, outreach, and educational purposes, targeting a broad community comprising
professional and citizen scientists, as well as the general public.
The public data are available in several formats and cover topics ranging from the origin and nature
of ultra-high-energy cosmic rays, multi-messenger studies, and searches for large and intermediate-scale
anisotropies in the arrival direction of cosmic rays.
The users can also exploit the physical concepts governing the development of extensive air showers
and assess the properties of high-energy hadronic interactions at an energy range beyond that achieved
at the Large Hadron Collider.
The datasets are updated yearly and comprise 10\% of the raw and processed data acquired with the
surface and fluorescence detectors used for our publications.
The datasets also include 100\% of the scaler data, which can be used for low-energy cosmic-ray studies
with primary energies of $10\:\mathrm{GeV}$ to a few $\mathrm{TeV}$, and the atmospheric monitoring data
that can be used to study several atmospheric phenomena.
Also, a catalog containing the 100 most energetic events detected during the Auger Phase I from
January 1, 2004, to December 31, 2020,~\cite{PierreAuger:2022qcg} was recently added.
At the date of writing, the last update is from March 2024, and new updates are foreseen throughout
the whole lifetime of the Observatory.
Finally, the Auger Open Data portal has a Contact link, through which a FAQ section and a helpdesk for
additional support are provided.

Below, a summary of the content of each section is given.

\subsection{Open Data archiving policy}

In recent years, the scientific community has witnessed a shift in the requirements set by science funders,
publishers, and governmental agencies.
The introduction of data management and stewardship plans for new experiments running on public funding
has become a norm~\cite{Wilkinson:2016myn}.
While these plans may not be the ultimate goals by themselves, they play a crucial role in ensuring the
reproducibility of previous scientific results even years after the data were collected or the end of
the data taking.
Moreover, these plans could lead to further scientific discoveries or innovative analysis methods.

In this spirit, the Pierre Auger Collaboration is committed to open science in all domains.
Its Open Data has an archiving data policy that adheres to the FAIR - Findable, Accessible, Interoperable,
and Reusable principles.
The Open Data are released under the CC BY-SA 4.0 International License~\cite{ccbysa}, and all
databases have a unique digital object identifier (DOI) provided by the general-purpose data
repository Zenodo~\cite{zenodo} that must be cited by all Open Data users.

\subsection{Datasets}

In the Auger Open Data portal, a detailed explanation of the Pierre Auger Observatory, its data
reconstruction and selection criteria, and the content of each dataset is provided.
The datasets are divided into three categories: \textit{Cosmic-Ray}, \textit{Scaler}, and
\textit{Atmospheric}.
Below, we provide a brief description of each dataset.

\paragraph{\textbf{The Cosmic-ray dataset}} contains 10\% of all Phase I events recorded by
the Pierre Auger Observatory that have passed high-level quality selection criteria used in our
analyses and publications.
It currently comprises 81,121 cosmic-ray showers, of which 25,086 events were measured with the surface
detector (SD) array of $1500\:\mathrm{m}$ spacing and 54,481 with the SD $750\:\mathrm{m}$ array;
the latter is a new addition to the available catalog.
From these, a subset of 3,348 events is called \textit{hybrid}, meaning that they were recorded
in coincidence with the fluorescence detector (FD) of the Pierre Auger Observatory, 197 of which
were detected with the High Elevation Auger Telescopes (HEAT), the low-energy extension of the FD.
For all events, pseudo-raw data is available in a JSON format, whereas high-level information is
accessible in a CSV format.
The events were reconstructed with the Auger \Offline Framework~\cite{Argiro:2007qg}, our
official software.
The high-level selection criteria for the events are described in~\cite{PierreAuger:2020yab}
and~\cite{PierreAuger:2014jss} for SD events with zenith angles below and above $60^{\circ}$,
respectively, and in~\cite{PierreAuger:2021hun} for the hybrid events.

\paragraph{\textbf{The Scaler dataset}} consists of more than $10^{15}$ events recorded with the
surface detector stations in the so-called ``scaler'' or ``particle-counting'' mode from March 2005 to
December 2020.
The scaler mode counts the number of times per second that a given surface detector registers an
energy deposit of about $15$ to $100\:\mathrm{MeV}$, which corresponds to the detection
of air-shower particles from low-energy cosmic rays of energies ranging from $10\:\mathrm{GeV}$ to a few
$\mathrm{TeV}$~\cite{PierreAuger:2011fsf}.
These low-level data are available as a CSV file.
They can be used to analyze the temporal behavior of the number of counts due to terrestrial and
extra-terrestrial phenomena or to study solar transient events like the Forbush decrease or
modulations induced by the solar cycle.

\paragraph{\textbf{The Atmospheric dataset}} contains information about the weather conditions
measured at 5 or 10-minute intervals at five different locations of the Pierre Auger
Observatory~\cite{PierreAuger:2012rhm}.
All data can be downloaded in a zipped folder containing 5 CSV files.
The weather.csv file contains processed weather information about the time, temperature, pressure,
density, and the averaged density in the previous two hours used, for instance, to calculate the
corrections of the energy estimator for the SD events.
The remaining five pseudo-raw files, also given in CSV format, contain the time, air temperature,
relative humidity, average wind speed, and barometric pressure taken by the five weather stations.\\

In addition, auxiliary data files are distributed containing the list of the positions of the SD stations
and the exposure of the SD $1500\:\mathrm{m}$ and SD $750\:\mathrm{m}$ arrays.
Similarly, the viewing directions of the fluorescence telescopes' pixels, and the parameters needed to
calculate the acceptance of the fluorescence detector are given.

\subsection{Visualize}
All events from the \textit{Cosmic-Ray} dataset can be visualized with the Event Browser available in the
\textit{Visualize} tab.
Some example events are immediately given at the top of the Event Browser.
These include the highest energy event in the dataset, the one with the highest number of SD triggered
stations, and a few hybrid events, among which are the so-called \textit{Golden Hybrid} events, i.e.,
the highest quality events used in our analyses, or the \textit{Stereo} events, which are seen by several
FD sites.
The user can also search for specific sets of events by applying cuts to the number of SD stations,
array type, energy, zenith angle, GPS time window range, or whether the event is hybrid.
Alternatively, a specific event can be selected by providing the browser with its ID.
After the selection is made, the events can be browsed in different tabs that can either give the event
description, energy, and zenith angle, or display the position in the SD array in which the event was
detected, and also check the traces of the triggered SD stations, its reconstructed lateral distribution
function (LDF), and, if the event is hybrid, the camera view of the triggered FD stations, and its profile
of deposited energy in the atmosphere.
The last tab prompts a 3-D animation of this event using the Unity engine~\cite{unity}, also widely
used for Outreach activities, particularly at the Auger International Masterclasses, which we will
address in the next section.

\subsection{Ultra-high-energy cosmic-ray catalog}
In addition to the \textit{Cosmic-Ray} dataset, the Pierre Auger Collaboration makes available its
collection of the 100 highest-energy events recorded between January 1, 2004, and December 31, 2020,
and published in~\cite{PierreAuger:2022qcg}, together with nine very energetic events used in the
energy calibration of the SD array.
As for the \textit{Cosmic-Ray} dataset, the pseudo-raw data can be downloaded in JSON format, and the
high-level information can be retrieved in a CSV file.
A full description of all data parameters for both file types is given.
Moreover, in the \textit{Catalog display}, it is possible to visualize each one of the events with the
Event Browser described above by clicking on the respective event ID.

\subsection{Analysis and Outreach}
Within the \textit{Analysis} and \textit{Outreach} tabs, several Jupiter Notebooks written in Python
are available.
These can be downloaded or run directly in the web browser via the Kaggle environment~\cite{kaggle}.
The notebooks can be easily modified to accommodate new plots or add further details to the
proposed analyses.

\paragraph{\textbf{In the Analysis tab,}} two Jupyter Notebook tutorials teach users how to read the CSV
and JSON files and produce basic histograms.
Detailed information about the content and meaning of each data field is also provided.
Also available are several notebooks containing simplified code snippets used to reproduce several analysis
results published by the Pierre Auger Collaboration.
These consider the anisotropy searches in the arrival directions of ultra-high-energy cosmic rays,
the energy spectrum, the estimation of the depth of the shower maximum, the measurement of
the proton-air cross-section, the energy calibration of the SD array, and the weather correction of
the energy estimation for SD events.

\begin{figure}[!h]
\centering
\includegraphics[width=14cm]{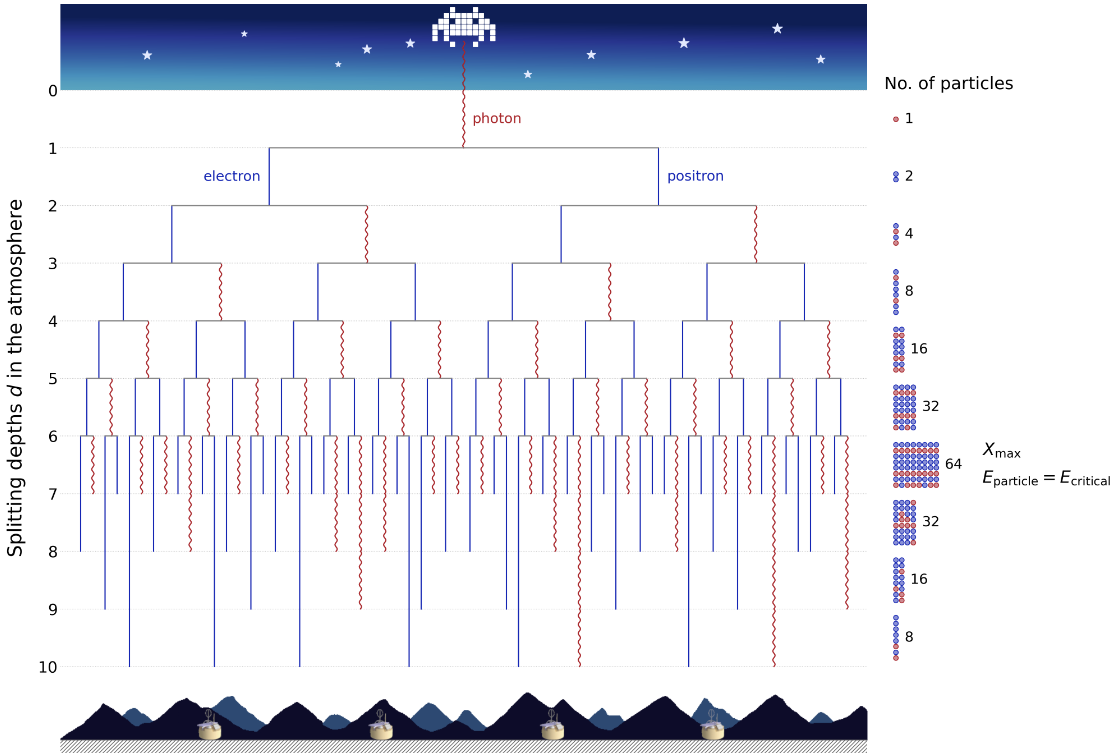}
\caption{From the \textit{Explore the shower development} Jupyter Notebook. Output of the animation code illustrating the development of a photon shower based on the Heitler model.}
\label{fig-1}
\end{figure}

\paragraph{\textbf{The Outreach tab}} follows a similar philosophy, but the provided Jupyter
Notebooks were adapted for a more general audience.
An introduction to cosmic rays and a description of the Pierre Auger Observatory translated into several languages
are given.
The provided Jupyter Notebooks are designed to be accessible to those with a basic understanding of Python
and data analysis.
These comprise a tutorial on how to read the CSV files and produce a set of basic histograms.
Then, users can explore the available datasets and plot data acquired by the SD and FD detectors from the
\textit{Cosmic-Ray} and also from the \textit{Scaler} and the \textit{Atmospheric} datasets.
Finally, one Jupyter Notebook explains the Heitler model, a toy-model that aims to describe the
development of the electromagnetic shower component in the atmosphere.
See Figure~\ref{fig-1} for an example photon-induced shower from the \textit{Explore the shower development} Jupyter
Notebook.

\section{Auger International Masterclasses}

The first International Masterclasses (IMC), promoted by the International Particle Physics Outreach Group
(IPPOG), were held across many European countries in 2005, the World Year of Physics.
The following year, institutes from the United States also joined the initiative.
Under the motto ``Become a scientist for one day,'' high-school students visit a research institution or
university, analyze experimental data, and have a chance to interact with scientists worldwide.
The most popular program offered by the IPPOG is called \textit{Hands-on Particle Physics}, where
students perform measurements with data from collider experiments at CERN.
Given the growing success of the International Masterclasses, the number of offered activities and their
scope have grown.
Currently, the IMC programs are held worldwide in 60 countries and are attended by more than 13,000
students.\\

In this spirit, and after a successful pilot program involving Italian and Portuguese students in 2022,
the Auger International Masterclasses~\cite{masterclasses} were proposed and included in the IMC
framework in 2023, benefiting from the IPPOG network of contacts of research institutions and schools.

\begin{figure}[!h]
\centering
\includegraphics[width=14cm]{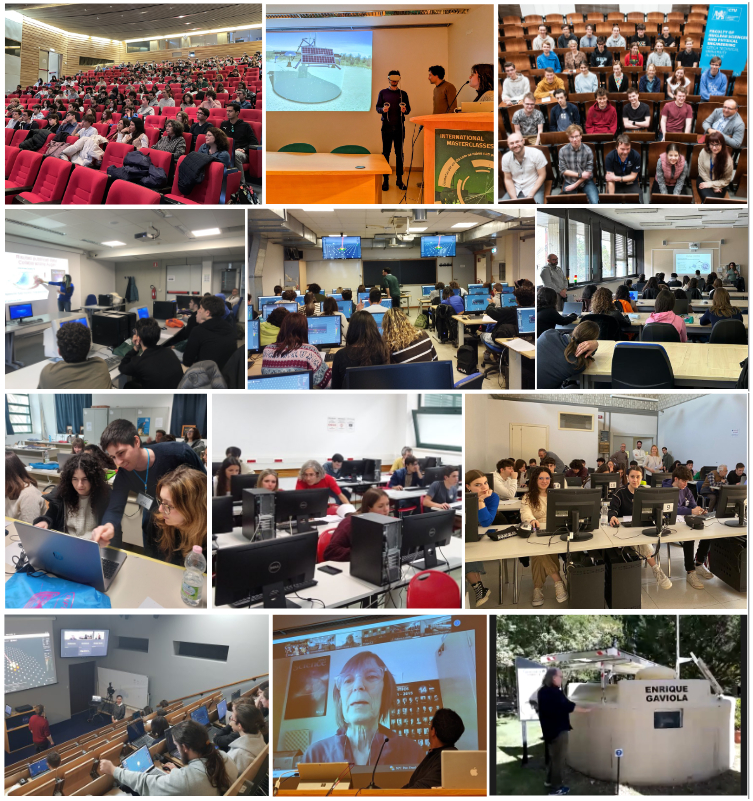}
\caption{Some photos of the Auger Masterclasses 2023 taken in several countries, and at the Pierre Auger Observatory.}
\label{fig-2}
\end{figure}

The debut of the Auger International Masterclasses was held in 2023 and successfully repeated in 2024.
The first Auger IMC were held on March 18, 24, and April 4 and were attended by more than 550 high-school
students at 12 research institutions from five countries on two continents, namely, the Czech Republic,
Italy, Portugal, and Romania in Europe, and Algeria in Africa.
Some photos of the several Auger International Masterclasses 2023 taken from several places around the
world, including at the Pierre Auger Observatory in Malarg\"{u}e, are shown in Figure~\ref{fig-2}.
In 2024, another 550 students from 10 countries from 4 continents, namely, Africa, Asia, Europe, and
North America, attended the Auger IMC.
This time, the activity comprised 16 research institutions, six of which did not belong to the Pierre
Auger Collaboration.
The participants were from Algeria, Kenya, Mexico, China, Japan, the Czech Republic, Italy, Hungary, Portugal, and Romania.

\subsection{How is it to be an astroparticle physicist for one day?}

The goal of the Auger International Masterclasses is to unravel the origin of the highest energetic
particles in the Universe.
For that, the participants are invited to analyze a set of Auger public events available at the Auger Open
Data portal under the assignment \textit{What is the origin of ultra-high-energy cosmic rays?}, and try
to reproduce the dipolar large-scale anisotropy observed in the arrival directions of the highest energy
cosmic rays detected by the Pierre Auger Collaboration, as published in~\cite{PierreAuger:2017pzq}.\\

A typical day at the Auger International Masterclasses starts with a morning session during which the
students have several lectures on astroparticle physics and how to detect cosmic rays.
The Pierre Auger Observatory and its detectors are explained as a requirement for the proposed hands-on
activity.
Finally, students will experience how it is to be in the Argentinian pampa or inside a surface detector
using virtual reality glasses.

After lunch, the students start the hands-on activity using independent datasets of 50 public events.
The students can work alone in groups of two people per desktop or laptop.
A printed guide explaining the activity is given, and tutors in the room can provide all the help and
guidance the students may need.
Typically, there is a tutor for ten students.

After the hands-on activity, which should last about two hours, all the students meet again and have a
video conference with scientists working at the Pierre Auger Observatory in Malarg\"{u}e.
During the video conference, the students discuss the results obtained, have a virtual tour of the
Observatory, and have a session of questions and answers with the Auger scientists.
Finally, using their cell phones, the students participate in a friendly competition in which they must
reply to a final quiz to test their knowledge acquired during this day.

\subsection{Hands-on: \textit{What is the origin of ultra-high-energy cosmic rays?}}

\begin{figure}[!h]
\centering
\includegraphics[width=12cm]{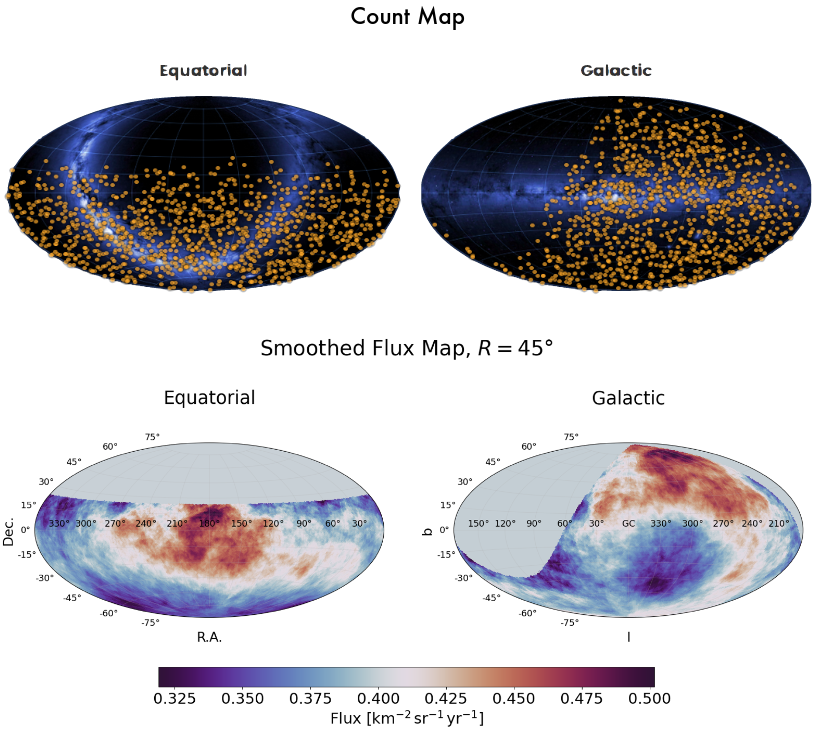}
\caption{The sky maps resulting from the first 2023 Auger Masterclass session, built with the arrival directions of 1130 events reconstructed by the students.}
\label{fig-3}
\end{figure}

During the hands-on activity, the students are asked to analyze independent datasets containing 50 public
events recorded with the SD stations and produce a sky map containing the reconstructed arrival direction of
the events that pass the selection criteria.
All events are available in the Auger Open Data portal and contain the traces of the SD stations and the
arrival times of the shower front at each station.
After the data analysis, the students upload their files with the reconstructed arrival directions of the
selected events on the Masterclasses web page, which combines the results from all students and runs the
\textit{UHECR sky} Jupyter Notebook to produce the final map of the arrival directions of ultra-high-energy
cosmic-ray events.
One example obtained in one of the Auger Masterclasses comprised of 1130 reconstructed events is shown
in Figure~\ref{fig-3}.
At the end, the students discuss their results and the concepts of sky-map coordinates, propagation effects,
statistical and systematic errors, and directional correlation and compare the obtained results about the
position of the maximum amplitude of the dipole with the position of the Milky Way and nearby galaxies.\\

\begin{figure}[!h]
\centering
\includegraphics[width=15cm]{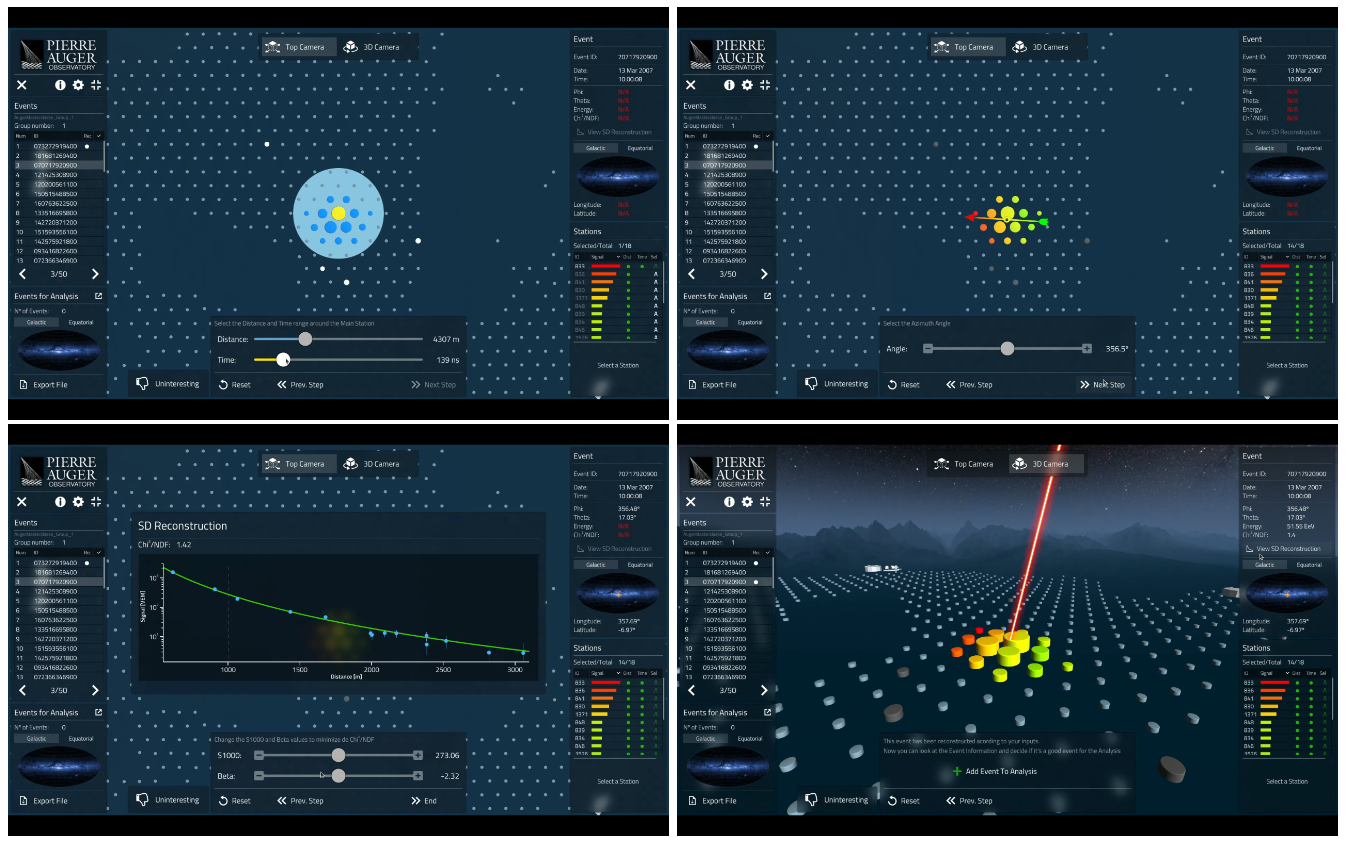}
\caption{The Auger Masterclass interface with the steps of the analysis of single events. From top left to bottom right, the steps are: 1) station selection, 2) azimuth (and zenith) angle reconstruction, 3) fit of a lateral distribution function, 4) acceptance or rejection of the event based on selection criteria.}
\label{fig-4}
\end{figure}

All the material required for the hands-on exercise, i.e., the 3-D Event Display made
in Unity~\cite{unity}, and the datasets can be downloaded from
\url{https://augermasterclasses.lip.pt/downloads}.
The event visualization is made with the 3-D Event Display.
The list of events to analyze is given in the left panel, whereas detailed information about the event
being analyzed is shown on the right panel.
The events can be rotated and zoomed in and out.
The SD stations participating in the event are colored according to the arrival time of the shower front
at the ground.
For the exercise, the position of the shower core for each event is given, from which the student must
then reconstruct the shower axis of the event, consisting of finding the zenith and azimuth angles of the
incoming shower, calculating its energy, and checking if the event passes all the selection criteria,
as described below.
In Figure~\ref{fig-4}, the several steps of the reconstruction process using the 3-D Event Display are illustrated.

\begin{enumerate}
 \item Station selection.
 \begin{itemize}
   \item The first step consists of selecting the SD station with the highest signal, called the hottest station, in the event by clicking on it.
  \item Next, two scrolls control the distance from the stations to that with the highest signal and the arrival time at a particular station. The student must move the two scrolls and select two opposite stations in the event so that the shower axis can be computed from the arrival time signal of the stations, assuming a plane shower front propagating at the speed of light.
 \end{itemize}
 \item Reconstruction of the the shower axis.
 \begin{itemize}
  \item The selected stations are colored according to the signal arrival time, and a direction must be visually adjusted to the color pattern to reconstruct the azimuth angle by choosing the first and the last station with a signal in the event.
 \end{itemize}
 \item Estimation the shower energy.
  \begin{itemize}
  \item Two scrolls control the interpolated signal at a distance of $1000\:\mathrm{m}$ from the shower axis, named $S\left(1000\right)$, and the slope $\beta$ of the lateral distribution function (LDF). The student should find the best reduced chi-square of the LDF fit.
 \end{itemize}
 \item Event selection.
 \begin{itemize}
  \item If the shower zenith and azimuth angles were calculated, and the LDF fit converges to a good value, then the event satisfies all selection criteria and should be saved. In this case, the arrival direction of the event is computed in equatorial and galactic coordinates and displayed on the sky map.
  \item Otherwise, if some of the previous steps failed, the event should be classified as uninteresting and not be considered for the analysis.
 \end{itemize}

\end{enumerate}

\section{Conclusions and Outlook}
The establishment of the Auger Open Data portal in February 2021 was the culmination of the Pierre Auger
Collaboration’s long and demanding commitment to making their data public, benefiting the scientific
community as part of the multi-messenger effort at different levels and for Outreach and Educational activities.
This work was only possible thanks to the help and support of many colleagues working on various topics within
the Pierre Auger Collaboration.
Moreover, releasing public data is a continuous task that will endure for the lifetime of the Collaboration, and
we hope to bear many fruits among the scientific and general community.
A strong synergy exists between the Open Data portal and the Auger International Masterclasses, integrated in
2023 as one of the IPPOG International Masterclass programs.
Given the two very successful Auger Masterclasses held in 2023 and 2024, this activity is foreseen to endure for
many other editions, spanning an increasing number of participants across many more countries worldwide.
The inclusion of further analyses using other Auger datasets, such as hybrid events, is also envisaged in the future.

\section*{Acknowledgments}
This work was co-funded by the European Union and supported by the Czech Ministry of
Education, Youth and Sports (Project No. FORTE – CZ.02.01.01/00/22\_008/0004632).

\end{document}